\documentstyle[preprint,aps]{revtex}
\tightenlines
\begin{document}
\draft
\title{
\bf{FINITE SIZE CORRECTION IN A DISORDERED SYSTEM -
A NEW DIVERGENCE }
}
\author{ Somendra M. Bhattacharjee\cite{eml1} and Sutapa
Mukherji\cite{eml2} }
\address{
Institute of Physics, Bhubaneswar 751 005, India
}
\date{\today}
\maketitle
\widetext
\begin{abstract}

We show that the amplitude of the finite size correction term
for the $n$th moment of the partition function, for randomly
interacting directed polymers, diverges (on the high temperature
side) as $(n_c - n)^{-r}$, as a critical moment $n_c$ is
approached.  The exponent $r$ is independent of temperature but
does depend on the effective dimensionality.  There is no such
divergence on the low temperature side ($n>n_c)$.
\end{abstract}

\narrowtext

The averages of thermodynamic quantities for quenched random
systems are often done through the replica trick which requires
an analytic continuation of the integer moments of the partition
function ${\overline{ Z^n}}$ to $n\rightarrow 0$ (See e.g.
\cite{mpv}), with the overbar representing the disorder
average. It is only recently that several issues related to this
continuation could be debated quantitatively because of the
results available for the problem of a directed polymer in a
random medium~\cite{comm}.  This was not feasible for other
models.  For example, the replica prediction of the nonexistence
of the higher order $(>3)$ cumulants of the free energy of 1+1
dimensional directed polymer problem has been challenged by
direct numerical computation of those\cite{karnpb,kbm}.  In
large dimensions, the behaviours of the moments for $n>1$ and
$n<1$ are found to be different\cite{bg}.  The possibility of
the existence of a singularity in $n$ has also been pointed
out\cite{ckder,mk93}.  All of these point finger to the analytic
continuation and interchange of the $n\rightarrow 0$ with the
thermodynamic limit.  In complex situations like the spin glass,
even the effective hamiltonian for the $n$th moment can admit of
many phases adding fuel to the fire\cite{sherr}.

In this paper we show an even more striking possibility in a
directed polymer system.  It is a {\it divergence} of the finite
size effect for the moments of the partition function in the
{\it high temperature phase} as a critical moment is approached.
This divergence is distinct from the known singularity, at the
thermodynamic transition point, in the finite size correction to
the free energy for disordered systems\cite{comm2} like spin
glasses\cite{tap}, the random energy model\cite{der80}, and
directed polymer in a random medium\cite{ckder}.  In contrast,
the divergence we report here is not necessarily at the bulk
transition point.

Our model involves two directed polymers which interact on
intersection at sites. The interaction energy is random with
zero mean.  The model is similar to those used for depinning
transitions of line like objects\cite{lei93}.  In the continuum
version, the random interaction model is known to exhibit a
disorder induced transition from a weak disorder (high
temperture) to a strong disorder (to be called the RANI phase)
phase\cite{ranil,ranie}.  For detailed understanding of the
phases, we study the model on a lattice using a real space
renormalization group approach.  Since real space
renormalization group can be implemented exactly on hierarchical
lattices
\cite{hier}, we consider the problem from the beginning on a
hierarchical lattice.  As usual, the procedure can be thought of
as an approximate real space renormalization group on a regular
lattice.

The lattice, as mentioned, is generated hierarchically by
replacing each bond at the $\mu$th generation by a ``diamond" of
$b$ branches (Fig 1).  Two directed polymers start at the bottom
of the lattice and meet at the upper end, without any
backtracking.  The model of interest is the site model where the
two polymers interact whenever they meet at a site, with a
random interaction energy $\epsilon$.  For simplicity, all the
directed paths are taken to be identical as far as the choice of
$\epsilon$ on these is concerned.  However, the energies along a
particular path are independent and uncorrelated, so that the
randomness is only in the longitudinal (special) direction and
not in the transverse directions (see fig 1). For this paper,
the only information needed about randomness is the behavior of
the moments of the weight $y = \exp (-\epsilon/T)$ ($T$ being
the temperature in units of the Boltzmann constant) which, for a
gaussian distribution of $\epsilon$, behaves as ${\overline{
y^m}} = {\overline{y}}^{m^2}$.  Other distributions have been
considered, though will not be elaborated upon here.

Let $Z_{\mu}$ be the partition function for a given realization
of randomness, and let $S_{\mu} = b^{L_{\mu} -1}$ be the number
of single chain configurations, at the $\mu$th generation.  Here
$L_{\mu}=2^{\mu}$ is the length of a directed polymer.  We
define ${\cal Z}_{\mu}(n) = {\overline{Z_{\mu}^n }} /
S_{\mu}^{2n}$, to factor out the free chain entropy, and call
${\cal Z}_{\mu}(n) $ the moments. We establish that, like the
random medium problem\cite{mk93}, for a given temperature, there
is a critical value $n=n_c({\overline{ y}})$ below which all
moments are in their high temperature phase, in the
thermodynamic limit of course.  In this limit, ${\cal
Z}_{\mu}(n)$ approaches a fixed point value for $n<n_c$,
whereas, for $n>n_c$, the moments diverge but with a finite
``free energy" density ${\sf f}_{\mu}(n) \equiv
(nL_{\mu})^{-1}\ln {\cal Z}_{\mu}(n) $(RANI phase).  The
approach to the thermodynamic limit can be written generically
as
\begin{equation}
g_{\mu}(n) = g(n) + {B_g(n)}\ {L_{\mu}^{-\psi_g}} + ...,
\label{eq:geng}
\end{equation}
where $g(n)$ is the thermodynamic limit ($\mu \rightarrow
\infty$), and $B_g(n)$
is the amplitude of the finite size correction.  We take
$g_{\mu}(n)$ for $n<n_c$ to be $[{\cal Z}_{\mu}(n)]^{1/n}$ , and
for $n>n_c$ to be ${\sf f}_{\mu}(n)$.  We will see that the
corrections are power laws in $L$.

One of the main results of the paper is the blowing up of
$B_z(n)$ as $ (n_c-n)^{-r}$, for $ n\rightarrow n_c-$, {\it a
feature whose existence was not even dreamed of before}.  No
such divergence occurs for $n>n_c$.  We study the variation of
$r$ with the various parameters of the problem and establish its
universality for a given distribution.  We also characterize the
behavior of ${\sf f}_{\mu}(n)$ for $n\rightarrow n_c +$.

For a given realization of disorder, the partition function can
be written as (see Fig
\ref{hier})
\begin{equation}
Z_{\mu+1}= b Z_{\mu}^{({\cal A})} y Z_{\mu}^{({\cal B})} +
b(b-1) S_{\mu}^4. \label{eq:zrec}
\end{equation}
The first term originates from the configurations that require
the two directed polymers to meet at $C$, while the second term
counts the ``no encounter" cases. There are no energy costs at
the two end points.  The moments of the partition function, from
Eq. \ref{eq:zrec}, are
\begin{equation}
{\cal Z}_{\mu+1}(n) = b^{-n} \sum_{m=0}^{n} P_{nm} {\cal
Z}_{\mu}^2(m) ,
\label{eq:rec1}
\end{equation}
where $P_{nm}={n \choose m} (b-1)^{n-m}{\overline{y^m}}$, with
the initial condition ${\cal Z}_0(n) = 1$ for all the moments
because there is no interaction in the zeroth generation (one
single bond).

The behaviour of the moments are determined by the stable fixed
points of the recursion relation as $\mu\rightarrow \infty$.
The first moment has no fixed point for ${\overline{ y}} > y_1
\equiv b^2/[4(b-1)]$.  For
${\sf y}_1\equiv{\overline{ y}} / y_1 < 1$, $ {\cal Z}_{\mu}(1)$
reaches a fixed point value ${\cal Z}^*(1) $ for large $\mu$.
For ${\sf y}_1 >1$, ${\sf f}_{\mu}(1)$ approaches a definite
limit.  For higher moments, the stable fixed points in extended
spaces can be determined with a high precision on a computer by
just solving a quadratic equation.  It is easy to see that, for
the stable fixed points of the first $n-1$ moments, there is
again a critical value $y_n$,
\[ y_n^{-1} = 4 b^{-2n} \sum_{m=0}^{n-1} P_{nm}[{\cal Z}^*(m)]^2
,\] so that for ${\sf y}_n\equiv{\overline{ y^n}}/y_n >1$ there
is no fixed point for the $n$th moment eventhough the lower
moments do have so.  The stable fixed point, if it is real, can
be written as
\begin{equation}
{\cal Z}^*(n) = b^n (2 {\overline{ y^n}})^{-1} \left[ 1 - \left
(1 - {\sf y}_n\right )^{1/2} \right ].\label{eq:brg}
\end{equation}
The property to be used later is that $2{\cal Z}^*(n) {
{\overline{ y^n}}} b^{-n} <1$.

{\it Numerical analysis:} By iterating the recursion relations
for given $b$ and ${\overline{ y}}$, we computed the moments of
the partition function with large (70 digit) accuracy using
Mathematica.  Iterations upto 100 generations are done, and for
checking, several cases with 300 generations are also
considered.  For a given ${\overline{ y}}$, there is a critical
value $n_c$ so that for $n<n_c$ the moments reach their fixed
point value, as shown in the lower inset of Fig \ref{delt}.
This $n_c$ depends on ${\overline{ y}}$ (i.e. temperature), $b$,
and the distribution. For $n< n_c$, the moments are in the high
temperature phase, as per Eq. \ref{eq:brg}, because for these
$y_n > {\overline{ y^n}}$. By the same token, for $n>n_c$, the
moments are in the low temperature phase.

To analyse the finite size data (finite generations) for $n <
n_c$, we adopt the following numerical procedure.  First,
construct the differences $\Delta_{\mu}(n)\equiv {\cal
Z}_{\mu+1}^{1/n}(n) - {\cal Z}_{\mu}^{1/n}(n)$, from Eq.
\ref{eq:geng}, as
\begin{equation}
\Delta_{\mu}(n)  =B(n) (1 - 2^{-\psi})\ L_{\mu}^{-\psi},
\end{equation}
omitting, for simplicity, the subscript $z$.  Hence, a log-log
plot would give $\psi$ and $B(n)$, provided $\mu$ is
sufficiently large. Such an analysis has been done for all the
moments for various ${\overline{ y }}$ and $b$. The sample plots
of Fig \ref{delt} clearly show that the exponent $\psi$ is
independent of $n$. It, however, depends on ${\overline{ y }}$.
(The upper inset of Fig \ref{delt} shows a few sample plots for
$[{\cal Z}_{\mu}(n)]^{1/n} -{\cal Z}^*(n)^{1/n} $ vs $\mu$
against $B(n) L_{\mu}^{-\psi}$ with the estimated values.) Fig
\ref{bn} shows the growth of the amplitude with
the moment indicating a divergence as the critical $n_c$ is
approached from below.  The location of $n_c$ and the exponent
$r$, as defined after Eq.
\ref{eq:geng}, can be determined by
choosing $r$ such that $B^{-1/r}(n)$ is a straightline with $n$.
The intercept gives $n_c$.

A similar analysis is done for ${\sf f}_{\mu}(n)$ for $n>n_c$.
The exponent $\psi=1$, and there is no divergence of the
amplitude near $n_c$.  The finite size correction, from Eq.
\ref{eq:rec1}, is $(\ln {\overline{ y^n}} - n \ln b) L^{-1}$.
In this case, the ``free energy", ${\sf f}(n)$ in the
thermodynamic limit vanishes in a singular fashion as $n_c$ is
approached .  Assuming ${\sf f}(n) \sim (n-n_c)^{\sigma}$,
$\sigma$ and $n_c$ can be estimated by linearizing the ${\sf
f}^{1/\sigma}(n)$ vs $n$ plot.

The estimates of $n_c$ from the two sides are consistent with
each other.  For $n<n_c$, see Fig \ref{bn}, fits over a wide
range in $n$ give the exponent $r$ as $0.71\pm .02$ and this
value is independent of temperature. We have checked this
insensitivity to temperature upto $\ln {\overline{ y}} = 0.005$
or $n_c =223.53$.  There is definitely a curvature near $n_c$,
and if only the last three or four points are taken, a lower
value for $r$ is obtained.  We, however, believe that this or
any other more elaborate procedure to estimate $r$ is not
warranted because the last few points are still far off from
$n_c$. We, therefore, tend to believe the obtained value of $r$
to be an upper limit like a mean field estimate.  The free
energy exponent $\sigma$ is found to be temperature dependent,
i.e., nonuniversal.  For example, for $b=4$, $\sigma$ changes
from $1.78\pm
.03$ at $\log {\overline{y}}= .04$ to $\sigma=1.9\pm .05$ at
$\log {\overline{y}}= .08$.  What really happens at $n_c$ eludes
us because it is not possible to hit an integer $n_c$
numerically.  It is plausible that $n\rightarrow n_c\pm$, and
$n=n_c$ are to be treated separately.

There is a strong dependence of the exponents on $b$ and the
distribution. For the gaussian distribution, the variation of
$r$ with $b$ is shown in Fig \ref{expr}.  This dependence might
be expected because as $b$ is changed the effective
dimensionality $2b=2^{d}$ changes.  Interestingly enough, the
exponent reaches a saturation as $b\rightarrow 2$, the minimum
value of $b$ for a transition.

{\it Linearized renormalization group analysis:} To rationalize
the results, we now analyze the linearized renormalization group
transformation.

For a given ${\overline{ y }}$, we linearize the recursion
relation, Eq. \ref{eq:rec1}, upto the moment for which the
stable fixed point is reached (i.e., $n < n_c$). Let us start
with the situation where the moments are close to the fixed
points, and define a column vector ${\bf z}$ of size $n_0$, the
integer part of $n_c$, with $z(n) = {\cal Z}_{\mu}(n) - {\cal
Z}^*(n)$ as the $n$th element.  The transformation matrix ${\bf
R}$ that takes the vector to a new one ${\bf z}^\prime = {\bf R}
{\bf z}$, is lower triangular with elements $ 2 b^{-n} P_{nm}
{\cal Z}^*(m)$, for $ m\leq n$, and zero otherwise. ${\bf R}$ is
of size $n_0
\times n_0$.

The eigenvalues of this matrix $\bf R$ are just the diagonal
elements $\lambda_n = 2 {\cal Z}^*(n) {\overline{ y^n}} b^{-n}$,
of which, in all cases, we find $\lambda_1$ to be the largest.
By Eq. \ref{eq:brg}, $\lambda_n < 1$ for all $n<n_c$.  (See Fig
\ref{eign} for the spectrum for a particular case.) If $\hat
{\bf e}_n$ is the $n$th eigenvector, then, after $\mu$
iterations, ${\bf z}^{(\mu)} = \sum_k A_k \lambda_k^{\mu} \
{\hat{\bf e}}_k,$ where $A_k$ is the projection of the starting
vector along ${\hat{\bf e}}_k$.  For large $\mu$, it follows
that the convergence to the fixed point for the $n$th moment is
as ${\cal B}_n L_n^{-\psi}$, with
\begin{equation}
\psi= -{\log_2 \lambda_1} = -\log_2 \left [1 - (1 -
{\sf y }_1)^{1/2}\right ],\label{eq:psi}
\end{equation}
for all $n$, and amplitude ${\cal B}_n = A_1 {\bf e}_{1,n}$,
where ${\bf e}_{1,n}$ is the $n$th element of $ {\hat{\bf
e}}_{1}$.

There is a universality in the exponent $\psi$ when considered
as a function of ${\sf y}_1={\overline{ y}}/y_1$.  The numerical
values for various temperatures, $b$, and distributions can be
made to collapse on this curve.  Incidentally, this is the same
exponent that a pure system would have, with attractive
interaction, if expressed as $y/y_c$ where $y_c$ is the
binding-unbinding transition.

The ratio $K_n$ of $n [{\cal Z}^*(n)]^{(n-1)/n} B_n/{\bf
e}_{1,n}$ is plotted against $n$ in Fig \ref{eign} for the same
situation.  This shows the proportionality of the two.  We have
ensured that this proportionality is maintained for all the
cases.  Hence, we infer that the divergence of the finite size
amplitude is really a consequence of the divergence of the
elements of the eigenvector for the largest eigenvalue.

How could the amplitude diverge? The eigenvalues are shown in
Fig \ref{eign} for one temperature.  In all cases studied, we
find the rise of the last few eigenvalues approaching from below
the first (largest) eigenvalue ($\lambda_1$).  This leads to an
increase in the components of the first eigenvector $\hat{\bf
e}$, whose $i$th component ${\bf e}_{1,i}$ is of the form
\[(\lambda_1 -
\lambda_i)^{-1} \left ( 1 + \sum_{j=2}^{i-1} \frac{R_{ij} R_{j1}}{
\Lambda_j }+  \sum_{j,k}
\frac{R_{ij}R_{jk}R_{k1}}{ \Lambda_j \Lambda_k} + ....\right),\]
where $\Lambda_p=\lambda_1-\lambda_p$. We, therefore, make the
following hypothesis.  In an analytic continuation in $n$, the
largest eigenvalue of the transformation operator (not
necessarily diagonal anymore) will be degenerate; the
eigenvalues of the modes at the end, near $n_c$, will come up
and merge with $\lambda_1$. It is this degeneracy that leads to
the singular behaviour and produces a nontrivial exponent $r$.

It is interesting to compare this situation with the problem of
a directed polymer in a random medium\cite{ckder,mk93,roux} and
the similar randomly interacting directed polymers but now
interacting on the bonds\cite{smhar}.  In these cases, unlike
Eq.
\ref{eq:zrec},  the temperature
does not enter the recursion relations explicitly but only
through the initial condition.  For this reason, linearizing
around the fixed point, with the associated eigenvalues and
vectors, will not produce any strong $n$ dependence, and hence
no divergence in the finite size correction.  In fact, an
attempt (not serendipity) to formulate a problem, where the
renormalization group transformations would contain all the
important information, led us to this particular site version of
the random interaction model.

In the context of the random medium directed polymer problem, in
1+1 dimensions, a scaling form has been proposed for the
moments, $\log {\overline{Z^n}} = n L f + g(nL^{\omega})$, where
$\omega$ is the free energy fluctuation exponent, and $f$ is the
thermodynamic free energy\cite{mk93}.  In this case, the whole
phase is the low temperature phase.  In our case, for $b>2$,
there is a phase transition which is reflected through a nonzero
$n_c$. The power law growth of the amplitude suggests a
different scaling form for $n<n_c$, namely $\log
{\overline{Z^n}} = n L f + g((n_c-n)L^{\Omega})$, where
$\Omega=\psi/r$.  It is tempting to speculate that this $\Omega$
is also the free energy fluctuation exponent. The question of
scaling on the other side however remains an open question.

To summarize, we have shown the existence of a diverging finite
size effect with a weakly universal exponent, at a critical
moment from the high temperature side - a scar left by the
disorder.  The thermodynamic limit on the high temperature side
is identical to the pure system. The growth of the amplitude
observed numerically can be a understood through a
renormalization group argument.  However, what controls these
exponents remains a puzzling issue.  A theory to understand the
universality (or its absence) of the exponent $r$ is lacking.
Full significance of the divergence is yet to be elucidated but
it cautions that an analytic continuation in $n$ has to be done
with proper care.

Partial support for this work has come from DST SP/S2/M-17/92.

\begin{figure}
\caption{ Construction of a hierarchical lattice. (a) This is
for b=2. Three generrations ($\mu$=0,1,2) are shown. For this
paper, identically marked points are taken to have the same
interaction energy $\epsilon$.  But, $\epsilon$ is random for
points with different markers. (b) A more general motif with
$2b$ bonds.}\label{hier}
\end{figure}
\begin{figure}
\caption{ Plot of $\log \Delta_{\mu} (n)$ vs $\log L_{\mu}$ for
$n=5$ (curve a), $n=20$ (curve b), and $n=26$ (curve c).  These
are for $b=4$ and $\log {\overline{y}} = 0.04$.  The straight
lines are the fits to these log-log plots.  The lower inset
shows the plot of ${\sf f}(n)$ vs $n$.  The transition is at
$n_c=26.6$.  The lower moments, $n<n_c$, are in the high
temperature phase while the higher ones are in the low
temperature RANI phase.The upper inset compares Eq. 1 with data
points for $n=5$(curve d) and $n=26$ (curve e) on a log
scale.}\label{delt}
\end{figure}
\begin{figure}
\caption{Plot of $[B(n)]^{-1/r}$ vs $n$ for $b=4$ and (a) $\log
{\overline{y}} = 0.065$, and $r=0.73$, (b) $\log {\overline{y}}
= 0.04$, and $r=.73$ (c) $\log {\overline{y}} = 0.03$, and
$r=0.72$ (d) $\log {\overline{y}} = 0.02$, and $r=0.72$. The
straightlines are the best fits through the points.}\label{bn}
\end{figure}
\begin{figure}
\caption{Dependence of the exponent $r$ on $b$.  The line
through the points is just a guide to the eye.}\label{expr}
\end{figure}
\begin{figure}
\caption{ (a)The 26 eigenvalues of ${\bf R}$ for $\log
{\overline{y}} = 0.04$ for which $n_c=26.6$.  Note the log scale
along y axis. (b) The horizontal straight line is the ratio
$K_n/K_1$. This plot is on a linear scale, shown on the
right.}\label{eign}
\end{figure}


\begin{references}
\bibitem{comm2} These singular corrections are obtained if the
averaged free energy is computed explicitly.  Unfortunately, the
replica method, though believed to be correct, has so far failed
to reproduce these singular terms of the free energy.
\end{references}
\end{document}